\documentclass[twocolumn]{article}
\usepackage[T1]{fontenc}
\usepackage{graphics}

\makeatletter

\newcommand{\LyX}{L\kern-.1667em\lower.25em\hbox{Y}\kern-.125emX\@}

\def\e{\mbox{${e^+e^-}$}} 

\makeatother

\begin{document}

\title{\textbf{\huge Multiplicity of different hadrons }\\
\textbf{\huge in \e, pp, and AA collisions}\huge }

\author{\textbf{H.J. Drescher, J. Aichelin, and K. Werner}\\
\\
\\
 SUBATECH \\
 Laboratoire de Physique Subatomique et des Technologies Associ\'{e}es\\
 UMR Universit\'{e} de Nantes, IN2P3/CNRS, Ecole des Mines de Nantes\\
 4, rue Alfred Kastler, F-44070 Nantes Cedex 03, France.\\
 and\\
 Nuclear Science Division, Lawrence Berkeley Laboratory, Berkeley, CA 94720,
USA \textit{}\\
}

\maketitle
\textbf{Abstract:} Employing the recently developed \textsc{neXus} model, we
compare the yields of different hadrons in ultra-relativistic collisions: electron-positron
(\e) annihilation at \( \sqrt{s}= \)91 GeV, proton-proton (\( pp \)) scattering
at \( \sqrt{s} \) =17 GeV and nucleus-nucleus (\( AA \)) collisions at \( \sqrt{s}= \)17
GeV (SPS) and \( \sqrt{s}= \)200 GeV (RHIC). Plotting the yields as a function
of the hadron masses, we find very surprising results: we observe that the spectra
are practically identical for \e at 91 GeV and central nucleus-nucleus reactions
at SPS and RHIC energies, whereas the spectrum for proton-proton scattering
is somewhat steeper. All have the form one expects if the particles were emitted
by a canonical system which is characterized by a temperature and chemical potentials.
These identical forms have, however, different origins: in \e and \( pp \)
the exponential shape it is due to the statistical behavior of string fragmentation,
which has absolutely nothing to do with thermalization, in \( AA \) it is caused
by phase space. The fact that \e and nuclear results agree is pure coincidence.
Surprisingly the results for \( pp \) and \e differ, although here the production
mechanism is identical. In pp collisions we see directly that the string energy
is very limited and hence the high mass baryons are suppressed. We conclude
that it is practically impossible to draw conclusions from hadronic yields about
the reaction mechanism.

\section{Introduction}

Studying hadron production in \e annihilation and pp scattering has already
a long tradition\cite{ep} \cite{pr}. In recent times particle production has
gained renewed interest because it has been argued that in heavy ion collisions
the enhanced yield of strange particles (as compared to pp) may serve as \char`\"{}smoking
gun'' for the conjectured creation of a quark gluon plasma (QGP)\cite{MR}.
Looking carefully at the presently available data of the reaction 158 AGeV Pb
+ Pb one observes indeed this fact \cite{WA} .

The formation of particles with a given mass depends always on the energy which
is available in the system which forms them. If this energy is not much higher
than the masses of the final state particles one expects that the phase space
of the strange particles is much smaller than the one of their lighter non-strange
counterparts and therefore they are less frequently produced. This is expected
to be the case in two body collisions in an expanding hadron gas with a temperature
of T=170 MeV as considered as a good description of the final state of the heavy
ion interaction. Contrary, if the energy of the forming system is much larger
than the hadron masses, the mass difference should not play an important role
and we expect to see (almost) the same yield. This scenario is expected if the
hadrons are produced by a decaying QGP.

In this situation it seems to be worthwhile to apply state-of-the -art models
in order to check the validity of the above mention conjecture. Here we use
the recently developed NEXUS model, whose results have been extensively compared
with most of the existing \e, \( pp \), \( \mu A \), \( pA \) and \( AA \)
data \cite{DHOPW}, \cite{DOPW}. For our study we compare the results of simulations
for 4 different reactions: \e at \( \sqrt{s} \)=91 GeV, \( pp \) at \( \sqrt{s} \)=17
GeV, central Pb-Pb collisions at \( \sqrt{s} \)=17 AGeV (SPS) and central Au-Au
collisions at \( \sqrt{s} \)=200 AGeV (RHIC).

\section{NE{\LARGE X}US}

How can one simulate in a realistic way hadron-hadron or even nucleus-nucleus
collisions? A sophisticated approach to such high energy interactions is the
so-called Gribov-Regge theory. This is an effective theory, which allows multiple
interactions to happen ``in parallel'', with phenomenological object called
``Pomerons'' representing elementary interactions. Using the general rules
of field theory, on may express cross sections in terms of a couple of parameters
characterizing the Pomeron. Interference terms are crucial, they assure the
unitarity of the theory. A big disadvantage of this approach is the fact that
cross sections and particle production are not calculated consistently: the
fact that energy needs to be shared between many Pomerons in case of multiple
scattering is well taken into account when calculating particle production (in
particular in Monte Carlo applications), but not for cross section calculations.
Another problem is the fact that at high energies one needs a consistent approach
to include both soft and hard processes. The latter ones are usually treated
in the parton model approach, which only allows to calculate inclusive cross
sections.

We recently presented a completely new approach \cite{DHOPW} for hadronic interactions
and the initial stage of nuclear collisions, which is able to solve several
of the above-mentioned problems. We provide a rigorous treatment of the multiple
scattering aspect, such that questions as energy conservation are clearly determined
by the rules of field theory, both for cross section and particle production
calculations. In both (!) cases, energy is properly shared between the different
interactions happening in parallel. This is the most important and new aspect
of our approach, which we consider a first necessary step to construct a consistent
model for high energy nuclear scattering. The elementary interaction is the
sum of the usual soft Pomeron and the so-called semi-hard Pomeron, where the
latter one may be obtained from perturbative QCD calculations (parton ladders).
To some extend, our approach provides a link between the Gribov-Regge approach
and the parton model, we call it ``Parton-based Gribov-Regge Theory''. Although
the main purpose of \textsc{neXus} is the simulation of nuclear collisions,
also electron-positron annihilation and deep inelastic lepton-nucleon scattering
have been intensively studied to make sure to have a consistent approach of
high energy scattering. 

The \textsc{neXus} simulation program is therefore available to simulate lepton-lepton,
lepton-hadron, hadron-hadron, hadron-nucleus, and finally nucleus-nucleus collision
at high beam energy. 

In electron-positron annihilation, starting from the virtual \( Z \), a time-like
cascade of parton develops, which are used to define a kinky string via identifying
partons and kinks. Particle production is the calculated using the standard
covariant string fragmentation procedures: the string breaks into string segments
vis quark-pair production, the segment representing the final hadrons and resonances.

In pp scattering, one has to deal with a complicated multiple scattering process,
but at the end the situation is quite similar to electron-positron annihilation
in the sense that parton configurations are translated into string language
in order to ``parameterize'' the theoretically unknown hadronization.

Concerning nucleus-nucleus collisions, we have to distinguish primary and secondary
interactions. Primary interactions are a straightforward generalization of proton-proton
scattering: multiple scattering -- parton production -- string formation. Whereas
in pp only few strings are formed and one may without problem consider independent
string fragmentation, this is not realistic for nuclear collisions. This fact
is take into account by so-called secondary interactions, going beyond \textsc{neXus}.
To get some idea what happens we employ a simple \textsc{droplet} model: based
on \textsc{neXus,} we calculate energy densities at a given proper time (say
at 1 fm/c). Spatial region with an energy density bigger than some critical
value are considered to ``quark matter'' and we treat the further development
macroscopically, assuming Bjorken-like longitudinal expansion and some transverse
expansion. Finally, when the energy density drops below the critical density,
the droplet decays into hadrons, purely via phase space. This approach proved
to provide a very satisfactory description of the CERN heavy ion reactions \cite{DOPW}.

\section{Results}

Before embarking to detailed comparisons and conclusions, we have to make sure
that the model describes reasonable well the observables we use here. Here we
are interested in the multiplicity in 4\( \pi  \) and therefore it is useful
to compare the Nexus results with experiments as well in 4\( \pi  \) \cite{pbm,sta,bar}. 
\begin{figure}
{\par\centering \resizebox*{!}{0.35\textheight}{\includegraphics{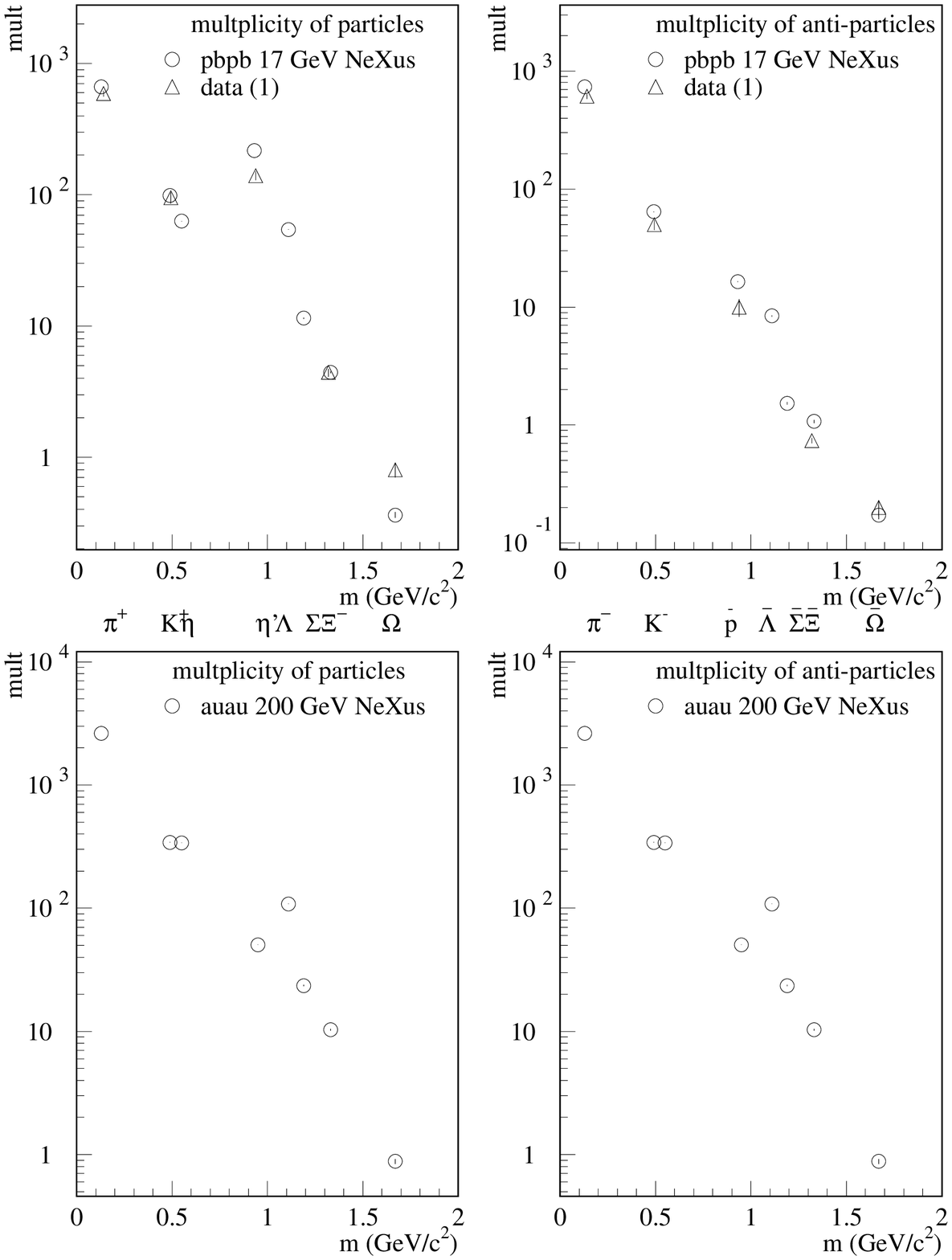}} \par}

\caption{Top row: The multiplicity per event of selected hadrons in 4\protect\( \pi \protect \)
as a function of their mass as compared to the experimental values which have
been extrapolated to 4\protect\( \pi \protect \). Bottom row: Predicted multiplicity
(after resonance decay for the RHIC experiment Au+Au at GeV. On the left hand
side we display mesons and baryons, on the right hand side mesons and anti-baryons.\label{johanna}}
\end{figure}
In fig. \ref{johanna}, we compare the 4\( \pi  \) Nexus results for particle
multiplicities (after decay of all not identified resonances) for central events
with a compilation of the experimental data. The \( \Omega  \) yield was obtained
by assuming that the ratio of \( \Xi  \) to \( \Omega  \) at mid-rapidity
obtained by WA97 is the same as the one for \( 4\pi  \). On the right hand
side, we display mesons and baryons, on the left hand side mesons and anti-baryons.
 The Nexus results are in good agreement with the extrapolated experimental
data over four orders of magnitude. In view of a similar good agreement for
\e and pp \cite{DHOPW} it is useful to use Nexus for the question at hand.
In fig. \ref{johanna}, we display at the bottom our predictions for central
collisions of Au + Au at \( \sqrt{s} \) = 200 AGeV (here as well non identified
resonances have decayed).

\begin{figure}
{\par\centering \resizebox*{!}{0.2\textheight}{\includegraphics{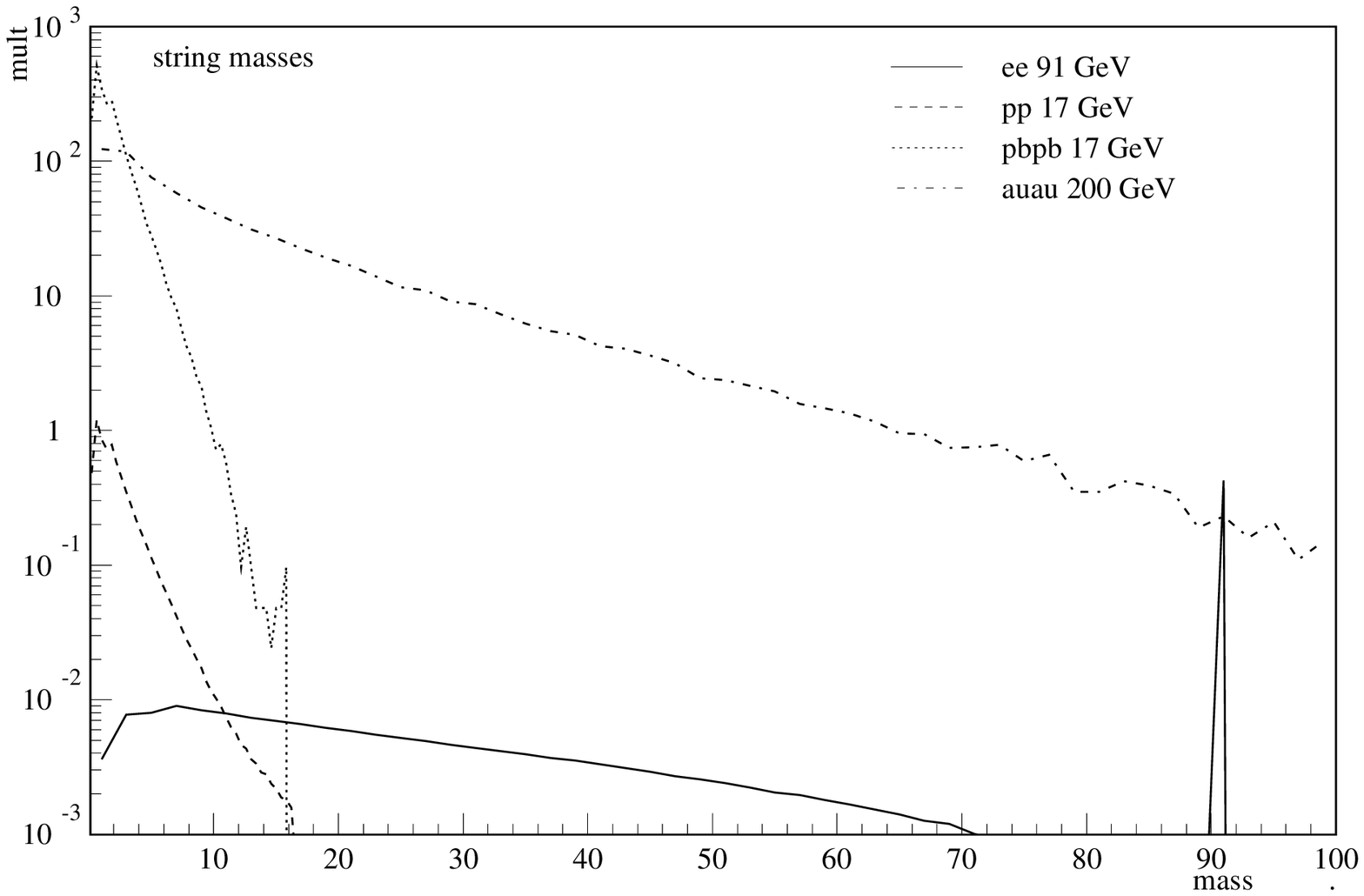}} \par}

\caption{Distribution of the string energies in \e at \protect\( \sqrt{s}=91GeV\protect \)
, pp and Pb+Pb at \protect\( \sqrt{s}\protect \) = 17 GeV and Au+Au at \protect\( \sqrt{s}\protect \)=200
GeV.\label{sqrts}}
\end{figure}

The important difference of pp and AA as compared to \e reactions is the distribution
of the energy W of the strings, i.e. the energy which is available for the production
of particles. For the 4 different reactions it is displayed in fig.\ref{sqrts}.
In the \e reactions at 91 GeV gluon bremsstrahlung rarely produces a quark-anti-quark
pair and hence the decaying string has almost always the full energy of 91 GeV.
In pp and AA reactions, on the contrary, only a small fraction of the initial
energy of the protons is available in the strings. The distribution of the string
energies for the reactions at \( \sqrt{s} \) = 17 GeV has a peak at around
1GeV and falls steeply off towards the maximal value of 17 GeV. The string contains
only a small fraction \( W=\sqrt{x_{1}x_{2}s} \) of the available energy \( \sqrt{s} \),
due to the fact that the light cone momentum fractions \( x_{1} \) and \( x_{2} \)
of the partons representing the string ends are effectively given by structure
functions which peak at low x. This peak value is small as compared to the threshold
for multi strange baryon production (due to strangeness conservation the threshold
for \( \Omega  \) is around 3 GeV). 
For RHIC energies the average string energy is already very large as compared
to the mass of stable baryons but still smaller than that for \e .  

\begin{figure}
{\par\centering \resizebox*{!}{0.5\textheight}{\includegraphics{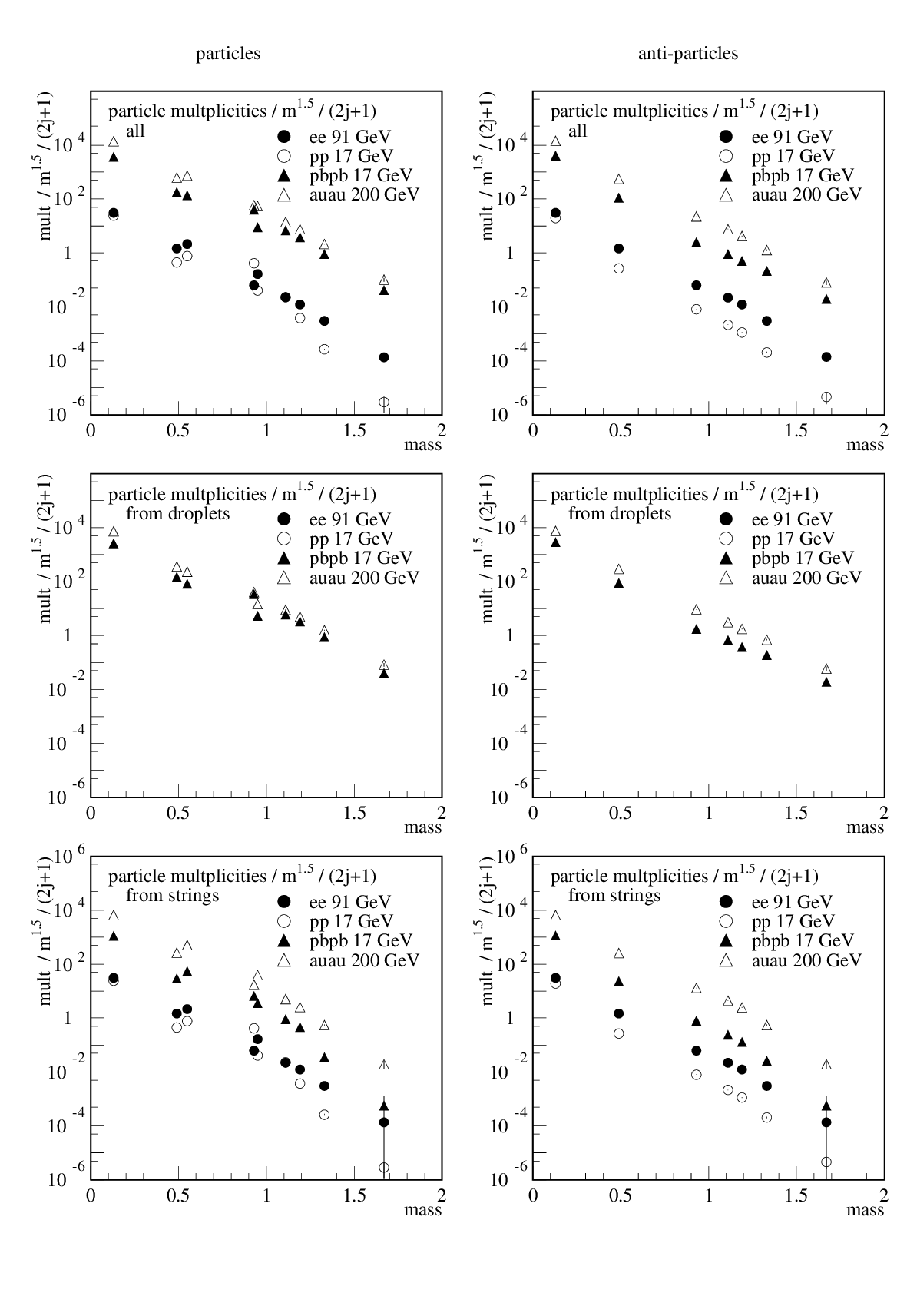}} \par}

\caption{Multiplicity/event in 4\protect\( \pi \protect \) as a function of the hadron
mass for the 4 investigated reactions. On the left hand side we display the
mesons and baryons, on the right hand side the mesons and anti-baryons. The
multiplicity has been divided by (2J+1) and \protect\( m^{3/2}\protect \) (see
text). Feeding from higher lying resonances has been excluded. In case of \protect\( AA\protect \)
collisions, we consider only those hadrons which are created directly from string
decay (not via droplets).\label{main}}
\end{figure}

\begin{figure}
{\par\centering \resizebox*{!}{0.2\textheight}{\includegraphics{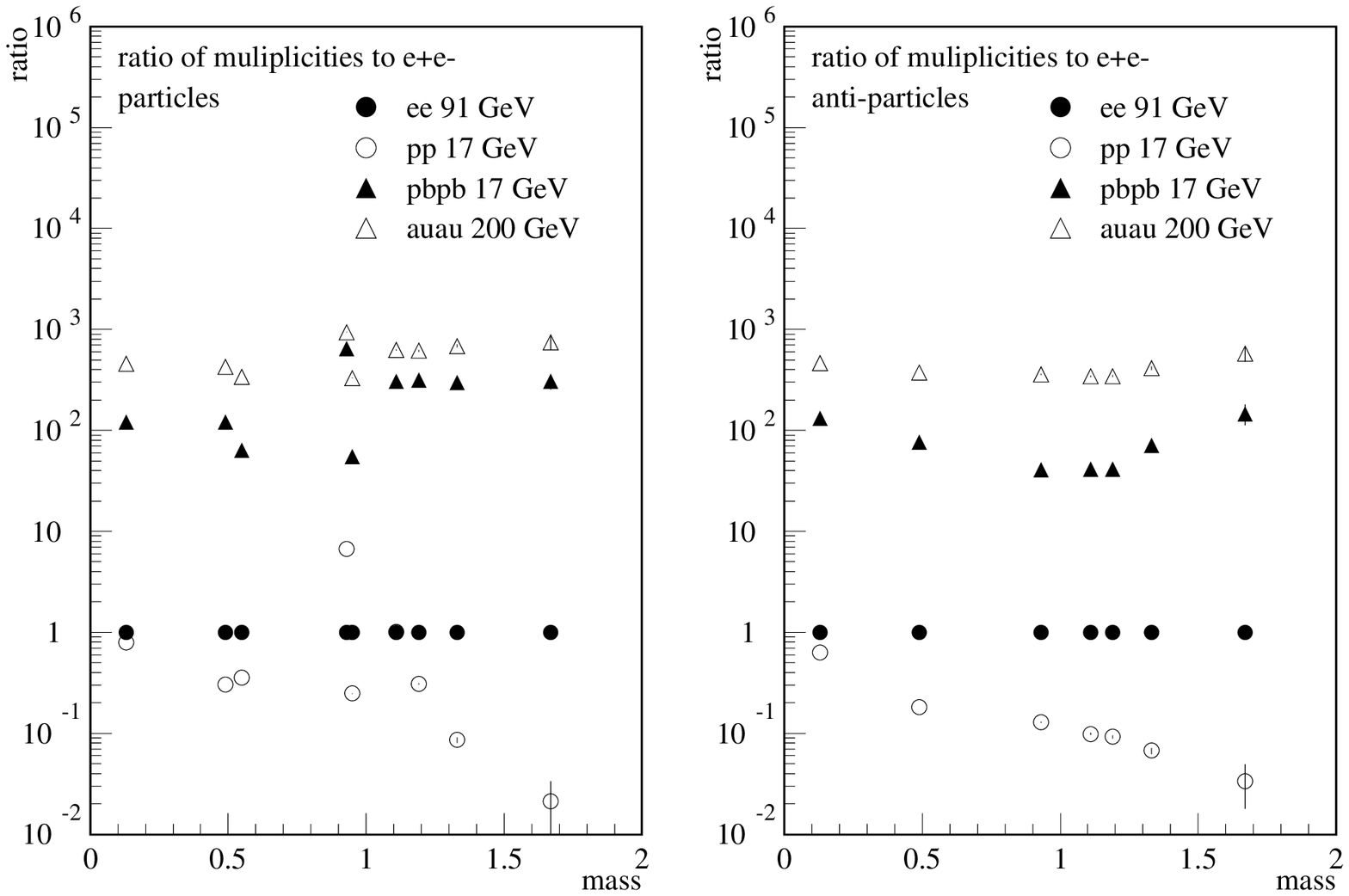}} \par}

\caption{Same as fig \ref{main}, but divided by the multiplicity observed in \e at
91 GeV.\label{quot}}
\end{figure}

Due to these distributions of string energies, we expect to see differences
in the yield of heavy as compared to light hadrons in the different reactions.
In fig.\ref{main}, we display the yields for the different reactions divided
by \( m^{(3/2)}(2J+1) \). In the case of \( AA \) collisions, only those detected
particles are considered which have been created directly from string decay
and not via droplets. 
In any case, we observe a roughly exponential drop of the yields as a function
of the mass: it is always more difficult to produce high mass particles compared
to low mass ones. The same exponential form of the mass yield is obtained in
this representation in canonical models which describe the mass yield in terms
of a temperature and chemical potentials \cite{pha}. Looking more closely we
observe, however, different slopes of theses exponential functions: \( pp \)
(17 GeV) and \( AA \) (17 GeV) show a similar behavior and both drop faster
than \e (91 GeV) and \( AA \) (200 GeV). In order to see the effects more clearly,
we divide all distributions by the \e results, see fig. \ref{quot}.
The \e (91 GeV) -- by definition -- and \( AA \) (200 GeV) show a horizontal
spectrum, whereas \( pp \) (17 GeV) and \( AA \) (17 GeV) show both a suppression
of heavy particles as compared to \e. At 17 GeV the string masses are small
and hence phase space reduces the production of heavy particles, whereas for
\e (91 GeV) and \( AA \) (200 GeV) there is no such limitation. 

\begin{figure}
{\par\centering \resizebox*{!}{0.2\textheight}{\includegraphics{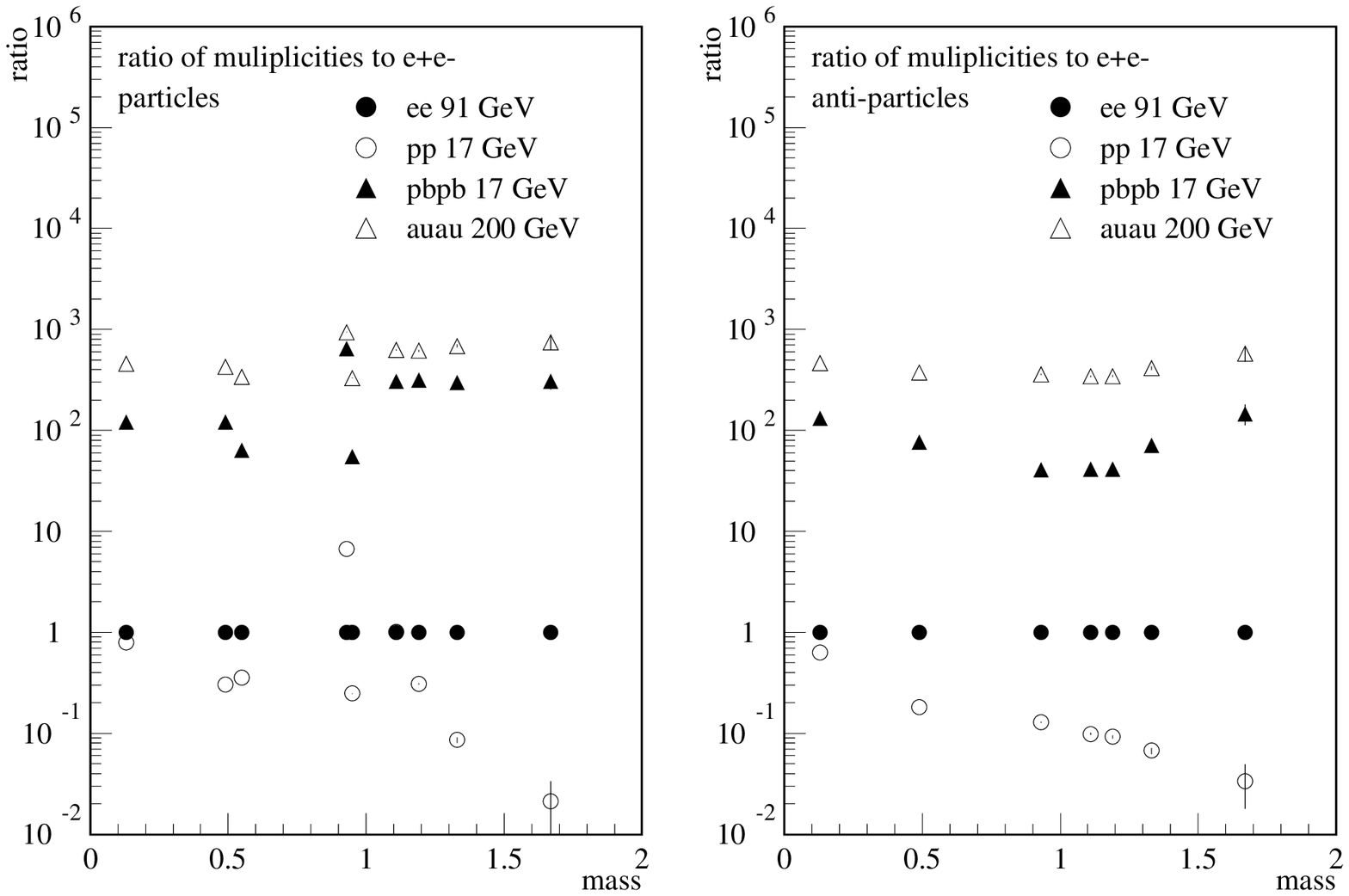}} \par}

\caption{Same as fig \ref{quot}, but showing the full result for \protect\( AA\protect \)
collisions, i.e. hadrons coming either from string decay or from droplets.\label{full}}
\end{figure}

So far we considered only \( AA \) collisions with hadrons coming directly
from strings. The majority of hadrons which are observed finally is, however,
produced from droplets. These droplets decay according to the microcanonical
phase space into baryons (the spin 3/2 decuplet and the spin 1/2 octet) and
mesons (pseudoscalar and vector octet) respecting all the conserved quantum
numbers of the droplet\cite{DOPW}. Their energy is usually large as compared
to the hadron masses. In fig. \ref{full}
we combine the hadrons produced by droplets and string decays and show the full
\( AA \) results as seen by the detectors, again compared to \( pp \) and
\e, and in all cases divided by the \e results. We observe that all the nuclear
results (\( AA \) (17 GeV) and \( AA \) (200 GeV)) are almost parallel, whereas
the \( pp \) spectrum drops much faster. If we take \( pp \) as a reference,
we might conclude: we observe indeed an enhancement of heavy (multi-strange)
hadrons for nuclear collisions but also for \e, however, in view of the above
discussion it is more appropriate to talk about a strangeness suppression due
to the limited energy in pp.  

\begin{figure}
{\par\centering \resizebox*{!}{0.2\textheight}{\includegraphics{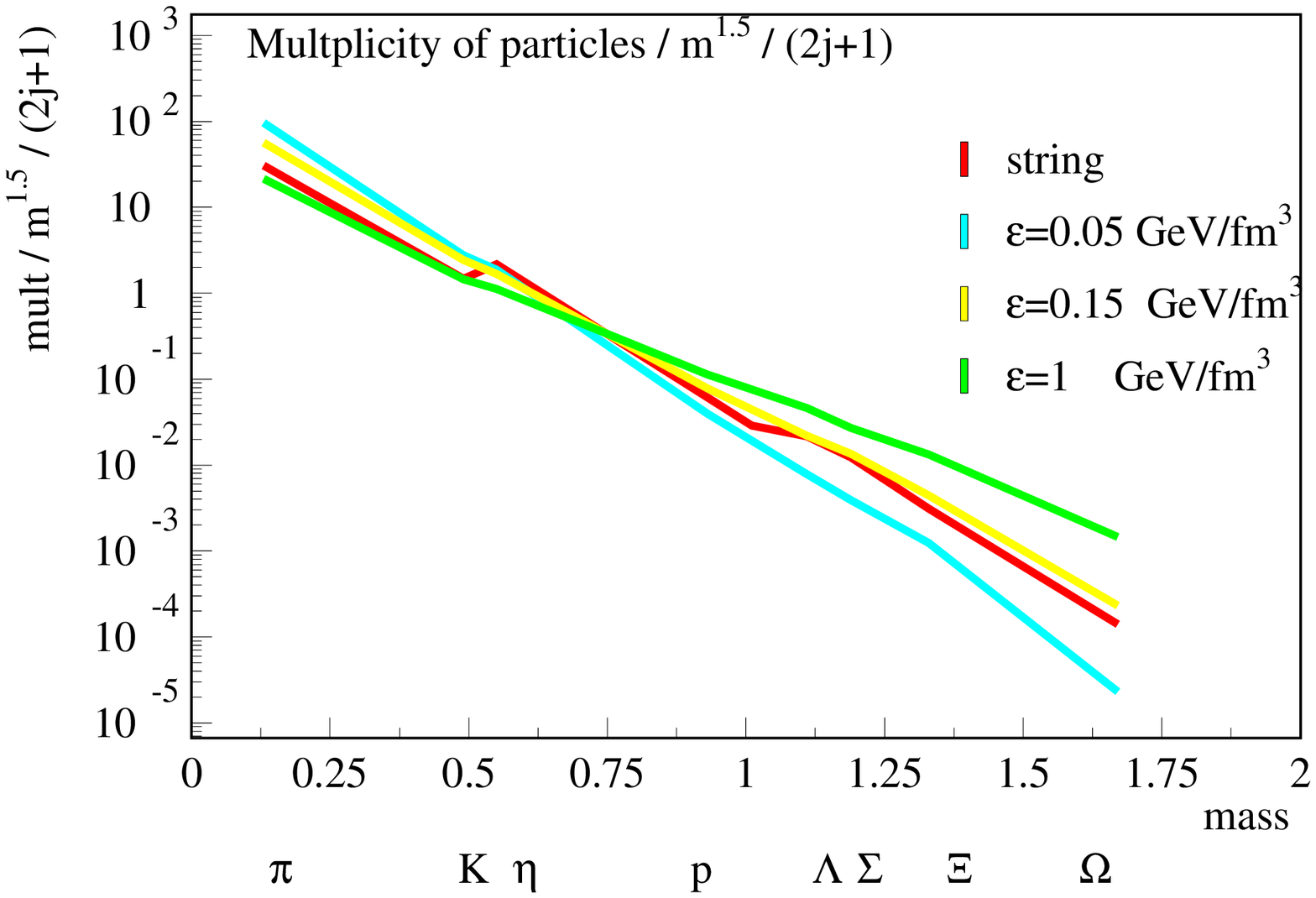}} \par}

\caption{Multiplicity/event in 4\protect\( \pi \protect \) as a function of the hadron
mass for \e
and droplet of 10 GeV assuming different freeze out energy densities \protect\( \epsilon \protect \).
\label{phas}}
\end{figure}

This is a very astonishing result in view of the fact that the origin of the
hadron production is very different in \e as compared to the hadronic interactions.
As said before, the decay of a string in \e is determined by the area law which
is close to a sequential decay in a one dimensional longitudinal phase space.
The droplet, on the contrary, disintegrates instantaneously and the different
decay channels are populated according to the three dimensional phase space.
Is it a accident that both mechanism give about the same hadron multiplicity
distribution? The answer is clearly yes as can be seen from fig. \ref{phas}.
 
The droplet decay pattern depends on the energy density \( \epsilon  \) where
the decay takes place. Smaller energy density implies larger volumes and therefore
decay channels in which many particles are produced are favored due to the additional
volume factor for each additional particle. If more particles are produced the
heavier ones are more suppressed, as can be seen. Now it happens to happen that
at energy densities of \( \epsilon _{^{\cong }}1GeV \), a number which can
be derived on theoretical grounds as well as by comparison with the data, \e and
hadronic interactions have about the same multiplicity distribution. This has
been already observed in thermal models but never understood in physical terms.
In the calculation presented here we have used \( \epsilon =0.15GeV \).

Comparing our completely microcanonical results with that of the canonical models
\cite{pbm,pha}, we see that both give about the same mass yield distribution.
The strangeness suppression factor which is used in \cite{pha} is a simple
parameterization for the fact that strange particles are heavier than their
non strange counterparts. It depends on the mass difference and hence is not
identical for all hadrons. This is in agreement with data.

In view of these results we can only draw the conclusion that at high energies
(i.e. if the average string energy is considerably larger than the masses of
the hadrons considered, as at RHIC energies) the functional form of the multiplicity
as a function of the hadron mass cannot serve as a \char`\"{}smoking gun\char`\"{}
for the plasma. A \e reaction, in which certainly no plasma but a string is
created, produces the same functional form. Hence the multiplicity distribution
cannot tell whether the strings, which are created in the AA reactions as well,
have interacted to form a QGP or whether they will disintegrate independently
without having it formed. Thus the quest for the QGP is more challenging than
thought.

Acknowledgments: We thank Johanna Stachel for fruitful discussions which have
initiated this work.

\end{document}